\documentclass[twocolumn,secnumarabic,amssymb, nobibnotes, aps, prd,superscriptaddress]{revtex4-2}
\usepackage{graphicx}
\usepackage{amsmath}
\usepackage{booktabs}

\begin{document}
	
\title{Machine learning enabled fast evaluation of dynamic aperture for storage ring accelerators}
\author{Jinyu Wan}%
\affiliation{Key Laboratory of Particle Acceleration Physics and Technology, Institute of High Energy Physics, Chinese Academy of Sciences, Beijing 100049, China}
\affiliation{University of Chinese Academy of Sciences, Beijing 100049, China}
\author{Yi Jiao}%
\email[Corresponding author:]{\\jiaoyi@ihep.ac.cn}
\affiliation{Key Laboratory of Particle Acceleration Physics and Technology, Institute of High Energy Physics, Chinese Academy of Sciences, Beijing 100049, China}
\affiliation{University of Chinese Academy of Sciences, Beijing 100049, China}
\date{\today}%

\begin{abstract}
For any storage ring-based large-scale scientific facility, one of the most important performance parameters is the dynamic aperture (DA), which measures the motion stability of charged particles in a global manner. To date, long-term tracking-based simulation is regarded as the most reliable method to calculate DA. However, numerical tracking may become a significant issue, especially when lots of candidate designs of a storage ring need to be evaluated. In this paper, we present a novel machine learning-based method, which can reduce the computation cost of DA tracking by approximately one order of magnitude, while keeping sufficiently high evaluation accuracy. Moreover, we demonstrate that this method is independent of concrete physical models of a storage ring. This method has the potential to be applied to similar problems of identifying irregular motions in other complex dynamical systems.
\end{abstract}
\maketitle

\setlength{\parskip}{0\baselineskip}
\section{Introduction}\label{section1}
Modern storage-ring accelerators, such as the synchrotron light sources \cite{ref1} and high-energy colliders \cite{ref2}, have played a key role in a variety of scientific fields like photon sciences \cite{ref3,ref4}, nuclear physics \cite{ref5,ref6}, material sciences \cite{ref7,ref8}, and high-energy physics \cite{ref9,ref10}. A storage ring can be regarded as a complex quasi-Hamiltonian system. The dynamical analysis for motions of charged particles in the storage ring remains an important and active issue in the accelerator field. The concept, dynamic aperture (DA, see details in Methods), is widely used to quantitatively measure the region of stable particle motions \cite{ref11}. DA is important for injection efficiency and beam lifetime \cite{ref12,ref13,ref14}, and is one of most important performance parameters of a storage ring. In the design of any storage ring, it is necessary to evaluate the DA and further optimize it (often needed) to meet injection and lifetime criterion.\par
In the last few decades, continuous efforts have been made to accurately predict the DA and to look insight the dynamics underlying the DA. Since early 1980s, analytical methods such as Lie algebraic methods \cite{ref15,ref16,ref17} have been transferred from the domain of nonlinear dynamical systems to DA evaluation and characterization. In the 1990s, numerical analysis methods, e.g., Lyapunov exponents \cite{ref18} and frequency map analysis \cite{ref19}, were introduced to characterize the stability of particle motions and reveal dangerous resonances limiting the DA. So far, a few analytical formulae (e.g., \cite{ref20,ref21,ref22,ref23,ref24}) and numerical tracking programs (e.g., \cite{ref25,ref26,ref27,ref28}) have been developed to evaluate the DA of a storage ring. Analytical approaches are fast, but have only limited accuracy in DA evaluation. Till now, long-term tracking-based simulation is regarded as the most reliable method to calculate DA. However, numerical tracking may become a significant issue in the cases a lot of candidate designs (or error models) of a storage ring need to be considered. Finding measures to effectively reduce the computing time for DA tracking will be very helpful.\par
To reduce the computational cost, machine learning (ML) techniques have recently inspired a surge of applications to DA evaluation, especially in the time-consuming DA optimization studies \cite{ref29,ref30,ref31,ref32,ref33}. These methods focus mainly on learning the map between the magnetic lattice settings of the storage ring (e.g., the strengths of magnets) and corresponding DA size. With a supervised ML model trained by existing data, the DA size can be predicted in a very short period of time and thus the DA optimization efficiency can be significantly improved. The ML model, however, does not always work well. This is because that the dynamics underlying the DA is very complex, and there is not explicit dependence between the DA and multi-kinds of related factors. Worse still, the available training data in hand is usually insufficient to cover a large enough volume in the variable space. As a result, the ML model often predicts accurately for a small variation range in the variable space and/or for a small number of control variables. Although this limitation can be mitigated to some degree by continuously retraining the ML model with new data samples (will take extra computing time) \cite{ref33}, one needs to always concern about the DA prediction accuracy when applying the ML model to a new storage ring design, whose lattice settings are away from the distribution of the existing training data.\par
In this paper, we propose a novel ML-based method of predicting DA. Considering the fact that judgment of particle loss or note over long-term tracking is the basis of the DA evaluation, we aim on training a ML model to learn the long-term motion stability from the initial trajectories of charged particles. In this way, the regression problem of predicting DA size of previous attempts can be reduced to a binary classification problem of predicting particle loss. Taking a fourth generation ring-based light source \cite{ref1} as an example, we will demonstrate that this method allows an accurate and fast DA evaluation process, and remains effective for different physical models of a storage ring.\par
\section{Methods}\label{sec:2}
\subsection{Conventional DA calculation}
\begin{figure*}[htbp] 
	\centering
	\includegraphics[scale=0.5]{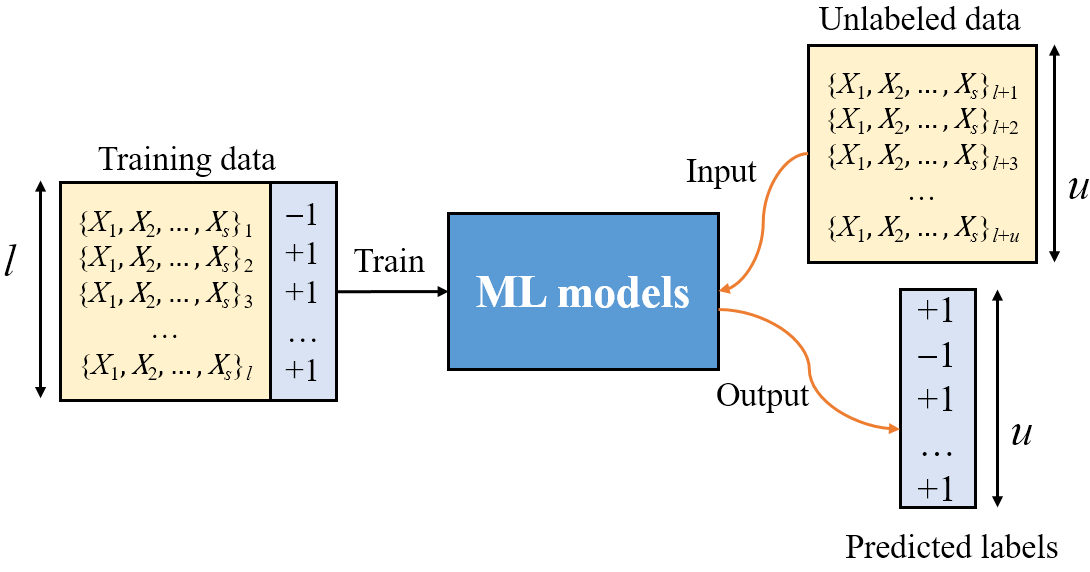}
	\caption{Schematic of how to predict particle loss from short-term particle motion trajectories. The surviving particles are assigned a label +1 and the lost particles are assigned a label $-1$. \emph{l} and \emph{u} are the numbers of labeled and unlabeled samples, respectively, where $\emph{l}\ll\emph{u}$. $\{\emph{X}_1, \emph{X}_2, ..., \emph{X}_\emph{s}\}_{i}$ represents successive particle trajectories in the phase space of s time intervals starting from the ith initial condition \{$\emph{X}_{0}$\}$_{i}$.}
	\label{fig1}
\end{figure*}
As a charged particle travels in a storage ring with a specific initial condition, the amplitude of its oscillation around the closed orbit \cite{ref34} may increase over time due to the presence of nonlinearities (e.g., from chromatic sextupoles), and in a worse scenario the amplitude exceeds a certain limit (e.g., physical apertures of the elements), leading to particle loss. The DA refers to the range of initial conditions of the particles whose trajectories stay within the specified limit for a sufficient number of turns. Strictly speaking, to represent the DA, one needs to count the initial conditions in 6D phase space, $[x, x', y, y', z, \delta$], where \emph{x} and \emph{y} ($x'$ and $y'$) are the horizontal and vertical displacements (angular deviations), $z$ and $\delta$ are longitudinal displacements and momentum deviation of the particle relative to a reference particle. However, for the sake of illustration, one generally searches for the largest initial amplitudes in transverse coordinate space, for which the subsequent trajectories are stable.\par
For a given lattice of the storage ring, a conventional way to calculate DA is to track particles over a range of initial conditions for a specified number of turns ($~10^3$ turns for lepton machines and up to $10^6$ turns for hadron ones). The initial conditions are usually uniformly sampled along multi-polar directions in transverse coordinate space. After checking the stability of subsequent trajectories for all initial conditions, the DA can be obtained by connecting the largest amplitudes for stable trajectories in different polar directions. Because of the heavy tracking simulation, a single evaluation of DA can take from minutes to days. The situation can be far worse in the case of DA optimization, in which one needs to evaluate the DAs for thousands of candidate magnetic lattices.\par
\subsection{ML-based scheme}
To evaluate DA in a faster manner, supervised ML methods are adopted in this study. A supervised learning task concerns to learn the correlations between data feature and label for a set of training data, and to predict labels of new data samples based on their data feature. Among various powerful ML methods, we select four widely-used supervised learning methods, which are support vector machine (SVM) \cite{ref36} based on statistical learning theory, random forest (RF) \cite{ref37} originated from the idea of ensemble learning, Gaussian process (GP) classification \cite{ref38} assuming the data subjected to a joint Gaussian distribution, and artificial neural network (ANN) \cite{ref39} simulating the neuronal mechanism of human brain. A semi-supervised extension of the SVM, transductive support vector machines (TSVM) \cite{ref40}, is also considered. These methods are tested and compared in Appendix A. It is found that RF shows better performance, and is therefore adopted in this study. \par 
In our DA prediction scheme, the initial conditions of particles are sampled in the same way as in conventional DA calculation. The difference is that particles are tracked for just a few turns (or only one turn). Among the initial conditions, only a small portion of them are randomly selected for long-term tracking. These long-term tracking results are used to train the ML model, where the first few turns' (or even the first turn's) trajectories are treated as feature X, and the corresponding stability of the trajectories are treated as label Y (+1 or -1 to represent stable or unstable). The schematic diagram of training is shown in Fig. 1. With the ML model, the stabilities of trajectories for the other initial conditions, which can be called unlabeled samples, are predicted as stable or unstable based on their short-term trajectories. After all initial conditions are assigned with labels (stable or unstable), the separation curve between the stable and unstable region in the initial condition space can be found, and the DA can be obtained. Apparently, since most of the particles are tracked for only a few turns rather than $10^3$ or even $10^6$ turns, the computation cost can be significantly reduced compared with a conventional computation process.\par
\section{Application to DA evaluation for a diffraction-limited storage ring}\label{sec:3}
\begin{figure*}[t] 
	\centering
	\includegraphics[scale=0.55]{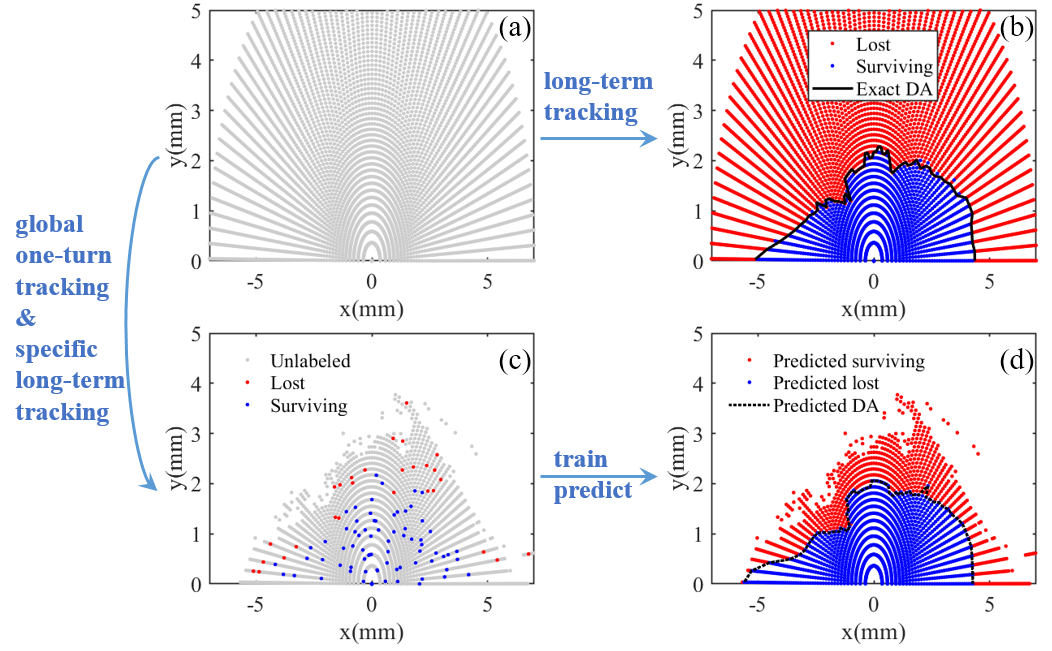}
	\caption{The comparison of the DA calculation with pure long-term particle tracking and the ML-based method. (a) shows the initial conditions of particles in the transverse space. (b) shows the long-term tracking results and the exact DA. (c) shows the particles that can survive the first-turn oscillation and the randomly selected initial conditions that are tracked for 1000 turns. (d) is the predicted long-term motion stability for unlabeled samples and the comparison with the exact DA. The red and blue points in (b) and (c) represent particles get lost and survive after 1000 turns tracking, respectively. While the red and blue points in (d) are particles being predicted to get lost or survive by the ML model, respectively.}
	\label{fig2}
\end{figure*}
In this section, taking a 4th-generation storage ring-based light source (4GLS) \cite{ref1}, the High Energy Photon Source (HEPS) \cite{ref41}, as an example, we will investigate the performance of the proposed ML-based DA prediction method. Nowadays, storage ring light sources have entered the era of their 4th-generation, which are usually called diffraction-limited storage ring. In a 4GLS, by using novel multi-bend achromats, the emittance of electron beam is pushed towards the diffraction limits of X-rays of interest to scientific community, so as to deliver photon beams of much higher quality than their predecessors and serve tens of users at the same time. The HEPS is such a 4GLS being built in China \cite{ref42}. After more than ten years evolution \cite{ref43}, the design and optimization of the storage ring has been finished \cite{ref44}. The HEPS storage ring comprises 48 seven-bend-achromats that are grouped in 24 super-periods, with a circumference of 1360.4 m and a natural emittance of ~34 pm. It provides 48 six meter long straight sections for insertion devices, producing X-ray radiation with brightness of above $1*10^22$ phs/mm2/mrad2/0.1\%BW. In the design of such a 4GLS, DA optimization is a challenging topic, because of the strong sextupoles required for chromaticity correction associated with the strong focusing required to reach an ultralow emittance design \cite{ref45}. With various efforts \cite{ref46,ref47,ref48,ref49,ref50}, the DA for the HEPS storage ring is only of the order of a millimeter, while meeting the requirement of on-axis injection \cite{ref51} to the storage ring. \par
In the following we use the conventional computation and the ML-based method to evaluate the on-momentum DA (related to the dynamical acceptance for a particle without any momentum deviation) for the bare lattice (without considering field and alignment errors), and then extend the studies to more practical models of the HEPS storage ring.\par
First, the on-momentum DA at the center of long straight section is computed with the program AT \cite{ref25}, basically following the conventional calculation process as mentioned above. In the calculation, a total of about 5,000 initial conditions of particles are sampled (see Fig. 2(a)) in the space of $[x, y]$, namely, with the initial conditions of $[x, x'=0, y, y'=0, z=0, \delta=0]$. A small difference from the calculation procedure mentioned in Sec. 2 is that the initial conditions with larger amplitudes are sampled with a slightly larger density (with a smaller step) to probe the DA size more precisely. The maximum amplitudes of \emph{x} and \emph{y} are set to 7 mm, to ensure the area covered by sampled initial conditions is sufficiently larger than the DA. Particles with different initial conditions are tracked for 1000 turns to see whether the particles survive or get lost, by checking whether the trajectories of particles exceed a limit of 1 m. The obtained DA is shown in Fig. 2(b). The DA calculation takes nearly half an hour on a personal computer. \par
Then we apply the proposed ML-based method to evaluate the on-momentum DA. For a fair comparison, the same initial conditions as in Fig. 2(a) are used, for which the particles are tracked for only one turn. Only the initial conditions of surviving particles are recorded in Fig. 2(c). It appears many particles with large initial amplitudes get lost within the first oscillation period in the ring, probably due to the presence of extremely strong nonlinearities. For those initial conditions left after one turn's tracking, the period-by-period coordinates are recorded and used as the feature of the samples. Then, from those initial conditions, 5\% of them are randomly selected for long-term tracking (1000 turns), and are assigned with a label (+1 or -1 to represent stable or unstable) based on whether particles survive or get lost after 1000 turns (marked in Fig. 2(c)). We used the selected samples to train the ML model. The hyper parameter settings of the training are introduced in Appendix A. With the trained ML model, the labels of the unlabeled samples (initial conditions with only the first turn's trajectories) are predicted to be lost or not. Then, the corresponding predicted DA is obtained by extracting the separation curve of the lost and surviving particles, with the results shown in Fig. 2(d). The whole process, including the time for training the ML model, takes nearly one minute on the same computer as used for the conventional computation.\par
\begin{figure*}
	\centering
	\includegraphics[scale=0.45]{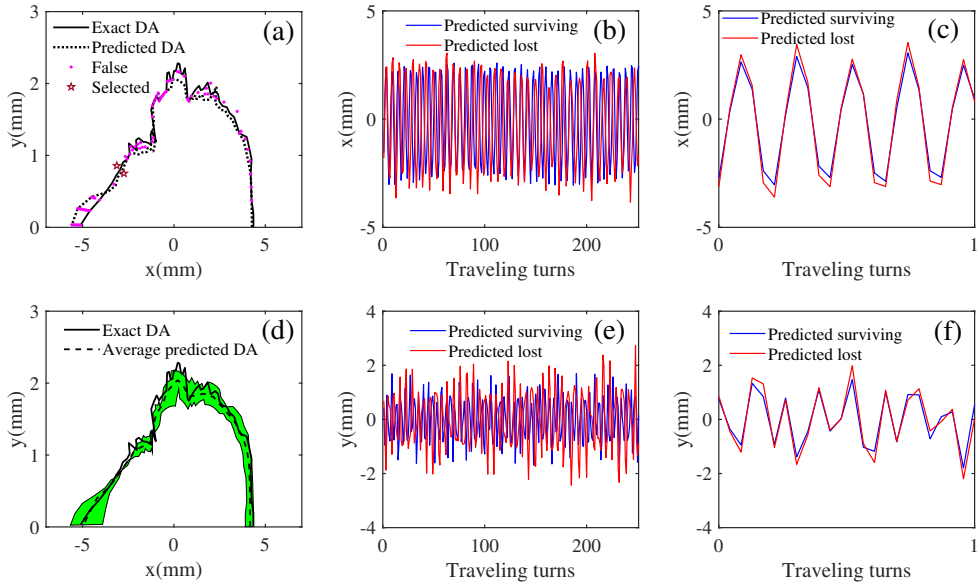}
	\caption{The comparison of the DAs obtained with pure long-term tracking and the ML-based method. (a) shows the comparison of the predicted DA and the exact DA in Fig. 2. The pink points in (a) represent false predictions of the ML model. (b) and (e) show the turn-by-turn evolution of the transverse amplitudes for two selected initial conditions marked in (a). The first-turn period-by-period trajectories for the two particles are also presented in (c) and (f). (d) shows average predicted DA among 100 repeated tests with different choice of training samples. The shadow in (d) are the maximum fluctuations of among the repeated tests. }
	\label{fig3}
\end{figure*}
For the convenience of comparison, the predicted DA and the DA obtained from pure long-term tracking (hereafter referred to as exact DA) are plotted together in Fig. 3(a). It appears the predicted DA accords with the exact DA pretty well. Some small details of the DA, e.g., a small reduction in vertical DA near $x=-1$mm possibly caused by resonance crossing, can be accurately detected by the ML model. Also, the false predictions occur only for the initial conditions close to the boundary of DA. Considering that the dynamics close to the DA boundary (transition between stable and unstable motions) is very complex and is really hard to predict, the ML-based method does pretty well in predicting DA. It is reasonable to deduce that the trained model captures most of the underlying correlations between the show-term trajectories of the motions and the corresponding long-terms stability.\par
\begin{figure}
	\centering
	\includegraphics[scale=0.4]{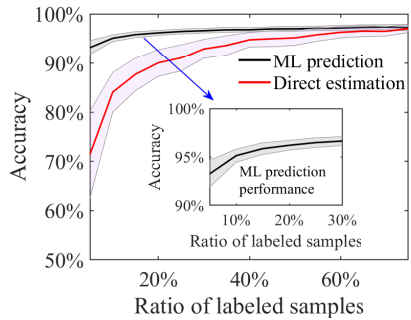}
	\caption{Accuracy of predicting DA for the HEPS by using different number of labeled samples. The red and blue solid lines represent average prediction accuracy among 100 repeated tests for the ML-based method and direct estimation, respectively. The shadow represents the standard deviation among the repeated tests. The direct estimation is implemented by connecting the outmost surviving labeled samples. The accuracy of direct estimation is slightly different from the definition of ML prediction accuracy, which is defined as the rate of the surviving and lost particles that are exactly separated by the estimated DA.}
	\label{fig4}
\end{figure}
As a further test, we compare the trajectories of particles from two close initial conditions (marked in Fig. 3(a)), which however, are within and outside the DA respectively and whose labels are accurately predicted by the ML model. Note that the two selected initial conditions are located in the region where false prediction rarely occurs, which can suggest the best performance of classification of the ML model rather than an average performance. The results are shown in Fig. 3(b) and (e). Although the difference is apparent between the two trajectories after tracking for more than 100 turns, there is just a little difference between the first turn's period-by-period trajectories (see Fig. 3(c) and (f)). With the ML model, the short-term trajectories are successfully differentiated and assigned a different label despite the very small difference between them. By contrast, previous experiences e.g., \cite{ref52} indicate that it is very difficult (if not impossible) to accurately judge the stability of motions based on such short-period trajectories.\par
Furthermore, to avoid the influence associated with the random choice of the training samples, 100 repeated ML-based DA predictions are done, with the results shown in Fig. 3(d). It shows that the maximum fluctuations and standard deviations of the predicted DAs are comparably small relative to the DA size, suggesting high stability of the ML-based method. The average predicted DA accords with the exact DA, with prediction accuracy of close to 95\%. The prediction accuracy is counted by calculating the rate of the number of accurately predicted initial conditions over the number of all unlabeled initial conditions.\par
\begin{figure*}
	\centering
	\includegraphics[scale=0.5]{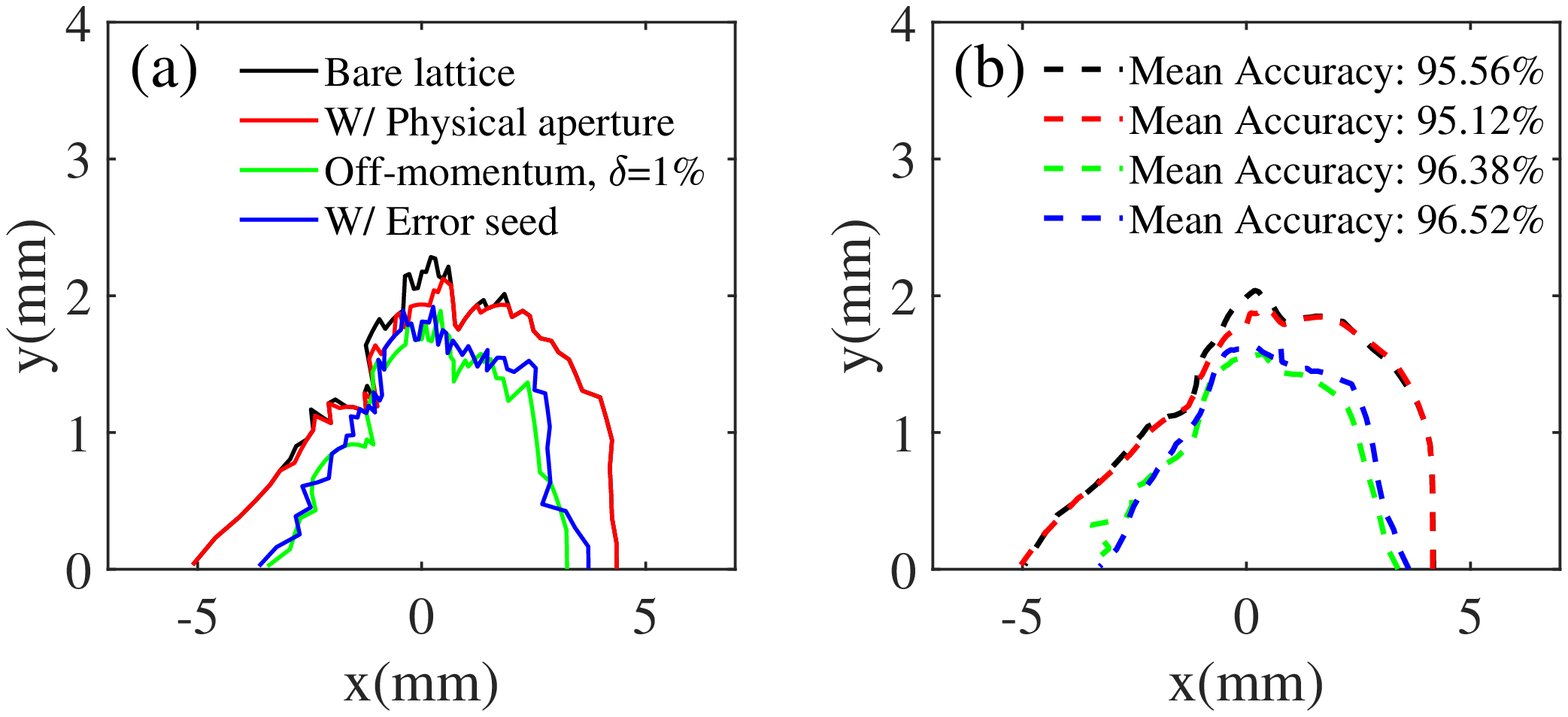}
	\caption{The DAs of the bare lattice, and the lattices in the presence of physical aperture of the HEPS, energy deviation, and magnetic strength error. (a) shows the exact DAs calculated by long-term particle tracking. (b) shows the average predicted DAs of the ML-based method among 100 repeated test by randomly selecting different labeled samples.}
	\label{fig5}
\end{figure*}
Based on the fact that the ML prediction performance is dependent on the quantity of training data, the labeled rate is scanned and the values of DA prediction accuracy are recorded. Here, the labeled rate refers to the ratio of the selected initial conditions to be used as training data over all initial conditions left after one-turn's tracking. The results are shown in Fig. 4. It appears that the prediction accuracy does increase when a larger set of training data is adopted. However, the increase in predication accuracy is no longer evident when the labeled rate is up to 10\%, and thereafter increases slowly and saturates at a value of approximately 96\%. From the figure, a labeled rate of 5\% appears sufficient and promising, which allows a good balance between the calculation cost and prediction accuracy. For a comparison, another alternative estimation method based on the same data is also studied, with the results also shown in Fig. 4. In this alternative method, the DA is estimated by directly connecting the outermost labeled surviving samples. One can image that if the labeled rate increases up to 100\%, the obtained result will be equal to the exact DA. The results appear that for this alternative method, to obtain high estimation accuracy of about 95\%, a much higher labeled rate than available with the ML-based method, i.e., at least 70\%, is needed.\par
Note that the DA computation results have a strong dependence on the physical model of a storage ring. Thus, we test the performance of the ML-based methods with modified lattices of the HEPS storage ring, where certain dynamical effects are included. Specifically, three cases are considered, off-momentum DA, DA in the presence of physical apertures of individual elements instead of a uniform limit of 1 m, and DA in the presence of magnetic field errors, respectively. Fig. 5 shows the corresponding exact DAs and the average predicted DAs among 100 repeated tests (with labeled rate of 5\%), respectively. It can be clearly seen the DA size changes as the model is modified. And, in all three cases, the DA prediction accuracy of the ML-based method remains as high as about 95\%. It appears that the proposed method is highly independent of the model. Actually, this is also demonstrated in another test, where the strengths of sextupoles and octupoles in the lattice are varied in a large range to search for a larger DA. \par

\section{Conclusion}\label{sec:5}
We have introduced a novel ML-based method of predicting DA, which can accurately predict the stability of massive particle motions in a storage ring, based on only a small amount of long-term particle tracking results together with a huge amount of short-term tracking results. We demonstrate that it is feasible to achieve high accuracy of approximately 95\% with only 10\% (or smaller) of long-term tracking results, at least in the presented study. Considering that the computing time for training ML model and short-term particle tracking is negligible compared with the long-term particle tracking, the proposed ML-based method would allow an order of magnitude faster DA evaluation, compared to conventional DA computation. Moreover, we show that this method is dependent of concrete magnetic lattices of the storage ring and the prediction accuracy of the proposed method remains high as the lattice is modified. This is inherent in the method based on the fact that it concerns only the correlations between the stability of motions and short-term trajectories, which is actually not related to the concrete models of the storage ring.\par
Extensive DA evaluations are required in many cases, for instances, testing the error tolerance of the storage ring design, and especially lattice and DA optimization in a preliminary design stage of a storage ring based-facility. In the latter case, due to the more efficient DA evaluation enabled by this novel method, it would be feasible to systematically explore ultimate performance of a specific lattice structure in a reasonable time, and implement thorough comparisons among different lattice structures to reach an final design with optimal balance between different design objectives.\par
Determination of motion stability is a fundamental issue not only for accelerator physics, but also among a large area of modern sciences, e.g., celestial mechanics and plasma confinement. Besides the applications to particle accelerators, we believe this model-independent ML-based method has the potential to be transferred to other similar complex dynamical systems. We have shared the open source code of a demo for this method on github (\url{https://github.com/wanjinyu/ML_DA}), and hope it is helpful to related studies in accelerator physics and other fields.\par

\section{Appendix A: Comparison of multiple ML methods}
\begin{figure*}
	\centering
	\includegraphics[scale=0.4]{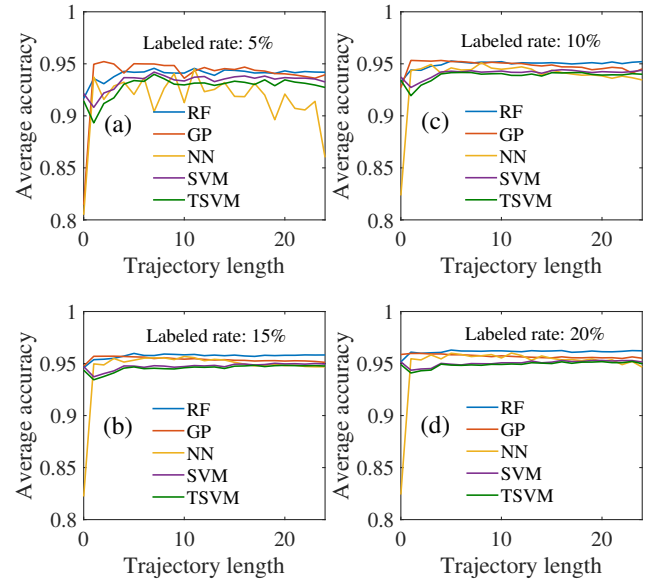}
	\caption{Comparison of prediction performance obtained with five ML methods. The trajectory length is defined as the number of super-periodic cells in the HEPS, which represents the length of the data feature in the learning task. A trajectory length of 0 represents only using initial conditions of particles to predict whether they are surviving or lost.}
	\label{fig6}
\end{figure*}
To explore better DA prediction performance, we compare four supervised ML methods, as well as a semi-supervised method that makes use of hidden information in the unlabeled data, which are RF, GPC, ANN, SVM, and TSVM, respectively. In this study, the following hyper parameters that are generally valid for most tested cases are chosen for these ML algorithms, which are decided based on extensive empirical comparisons. For RF, the number of decision trees is set to be 100. For GPC, the radial basis function is used as its kernel function, and the maximum iteration number for approximating the posterior during predict it set to be 100. For ANN, a network structure of only one hidden layer with 128 neurons connected is adopted. The maximum number of training iterations for ANN is set to be 3000. The penalty coefficient for SVM and TVSM is set to be 1.5 and a radial basis function is used as kernel function. The unlabeled samples are assigned a weight factor of 0.001 that exponentially increases in the training of TSVM. The random state of RF and GPC is set to be 0 to ensure that only the choice of training samples can affect the results among the repeated tests. The training of RF, GPC, SVM and TSVM is implemented on an open source ML library Scikit-Learn \cite{ref53}, and ANN is trained on another open source ML library tensorflow \cite{ref54}.\par
Besides the comparison of using different number of training samples, we also investigate the influence of length of particle trajectories that is the data feature on the prediction accuracy. Fig. 6 shows that for RF, the prediction accuracy can be improved by using longer motion trajectories when the trajectory length is less than 10, while further increase can hardly improve the accuracy. It is also observed that the prediction accuracy of some ML methods may even decrease when more motion information, i.e., longer motion trajectories in the phase space, is provided. This may be because of the sensitivity of their hyper parameter settings to specific learning tasks. To obtain the best performance, their hyper parameter settings may need to be optimized when the learning task is slightly changed, especially when the length of data feature vector is changed. Nevertheless, with a labeled rate up to 10\%, all tested ML methods imply similar prediction performance. Among the tested methods, the RF shows better stability of the prediction performance, and slight higher prediction accuracy when motion trajectories over more than 15 super-periodic cells in the HEPS are available. Thus the RF is adopted in this study to investigate the nonlinear dynamics of storage ring accelerators.

\section{Acknowledgement}
The authors thank Yongjun Li for nice discussion and suggestions. This work is supported by National Natural Science Foundation of China (No. 11922512), Youth Innovation Promotion Association of Chinese Academy of Sciences (No. Y201904) and National Key R\&D Program of China (No. 2016YFA0401900). 

\end{document}